\documentclass[sn-mathphys-num]{sn-jnl}

\usepackage{graphicx}%
\usepackage{multirow}%
\usepackage{amsmath,amssymb,amsfonts}%
\usepackage{amsthm}%
\usepackage{mathrsfs}%
\usepackage[title]{appendix}%
\usepackage{xcolor}%
\usepackage{textcomp}%
\usepackage{manyfoot}%
\usepackage{booktabs}%
\usepackage{algorithm}%
\usepackage{algorithmicx}%
\usepackage{algpseudocode}%
\usepackage{listings}%
\usepackage{optidef} 

\theoremstyle{thmstyleone}%
\theoremstyle{thmstyletwo}%

\theoremstyle{thmstylethree}%

\begin{document}

\title[Mechanistic interplay between information spreading and opinion polarization]{Mechanistic interplay between information spreading and opinion polarization}


\author*[1,2]{\fnm{Kleber} \sur{A. Oliveira}}\email{Kleber.AndradeOliveira@mtu.ie}

\author[1,3]{\fnm{Henrique} \sur{F. de Arruda}}

\author[1,4,5]{\fnm{Yamir} \sur{Moreno}}

\affil*[1]{\orgname{CENTAI Institute}, \orgaddress{\city{Turin}, \country{Italy}}}

\affil[2]{\orgdiv{Department of Mathematics}, \orgname{Munster Technological University}, \orgaddress{\city{Cork}, \country{Ireland}}}

\affil[3]{\orgdiv{Geography and Geoinformation Science, College of Science}, \orgname{George Mason University}, \orgaddress{\city{Fairfax, VA}, \country{USA}}}

\affil[4]{\orgdiv{Institute for Biocomputation and Physics of Complex Systems (BIFI)}, \orgname{University of Zaragoza}, \orgaddress{\city{Zaragoza}, \country{Spain}}}

\affil[5]{\orgdiv{Department of Theoretical Physics, Faculty of Sciences}, \orgname{University of Zaragoza}, \orgaddress{\city{Zaragoza}, \country{Spain}}}


\abstract{We investigate how information-spreading mechanisms affect opinion dynamics and vice-versa via an agent-based simulation on adaptive social networks. First, we characterize the impact of reposting on user behavior with limited memory, a feature that introduces novel system states. Then, we build an experiment mimicking information-limiting environments seen on social media platforms and study how the model parameters can determine the configuration of opinions. In this scenario, different posting behaviors may sustain polarization or revert it. We further show the adaptability of the model by calibrating it to reproduce the statistical organization of information cascades as seen empirically in a microblogging social media platform.} 




\maketitle

\section{Introduction}\label{sec1:intro}

Through their influential book, Jamieson and Cappella~\cite{jamieson2008echo} popularized the term `echo chamber' as ``a bounded, enclosed media space that has the potential to both magnify the messages delivered within it and insulate them from rebuttal''. More than a decade later, the availability of social media platform data propelled a whole field of intense research, permeated with debates on several levels of how to conceptualize~\cite{Nguyen_2020}, characterize~\cite{terren2021echo} and measure~\cite{cota2019quantifying} echo chambers.

Due to the frequent interchange with the term ``filter bubble''~\cite{ross2022echo}, some researchers started to refer to this class of phenomena as information-limiting environments~\cite{kitchens2020understanding}. Such environments, in turn, can be ascribed as one of the aspects of the larger research problem of online misinformation. The urge to build systematic scientific knowledge on the misinformation problem stems from its large-scale consequences on societal decision-making~\cite{bakcoleman2021stewardship},
down to the very fabric of democracy~\cite{lorenz-spreen2023democracy}.

However, building a scientific understanding of how online information affects its consumers is particularly difficult. This influence occurs through an intricate, multilayered combination of spontaneous user interaction and platform control, which is in the form of constantly updated features that are rarely disclosed to the general public or researchers. There is an increasing need to formulate these systems mechanistically so that one may decouple the effects of platform intervention and substantiate how responsible policies should be designed.

An important line of investigation of this problem is concerned with developing computational models to explain opinion dynamics~\cite{noorazar2020opiniondynamics}. Such models typically feature agents with a numerical representation of an opinion and are designed to capture behaviors emerging from agent interactions~\cite{deffuant2000mixing,hk2002opinion,galam2002minority}. In many of these, one is worried about system states of opinions highly concentrated on two opposing sides so as to depict ideological polarization~\cite{kubin2021polarization}.

Among opinion dynamics models, a number of them implemented rewiring mechanisms leading to adaptive networks under agents' interactions~\cite{holme2006nonequilibrium, benatti2020opinion, baumann2020modeling}. This has been identified as a mechanism for the emergence of mesoscale modular structures within the networks, which are often associated with echo chambers~\cite{sasahara2021social}. It has also been shown analytically that homophily-biased rewiring can divide networks into polarized communities~\cite{blex2022homophilic}.

In this context, the model by de Arruda~\emph{et al.}~\cite{de2022modelling} introduces a proxy for a recommendation algorithm on top of influencing and rewiring rules and is able to fragment the network into polarized communities even with non-biased rewiring. It has been shown by Valensise~\emph{et al.}~\cite{valensise2023drivers} to fit empirical data from the popular microblogging platform now known as X (previously Twitter\footnote{For the remainder of the paper, we refer to the platform as Twitter instead of X. This choice also reflects the fact that many datasets from the time it was named Twitter remain available. On the other hand, X's policy change now restricts data collection, rendering it difficult to study the most up-to-date data.}) better when compared to other similar models on polarized contexts resembling echo chambers.

Notwithstanding, a key ingredient to characterize information-limiting environments is how the information spreads, which is crucially related to the physics of social media systems \cite{ciampaglia2015attention}. Weng~\emph{et al.}~\cite{weng2012limited} discovered that on Twitter, the variety of content an individual pays attention to does not increase with the variety of content in the system, meaning there is a limit to each person's attention.

This limit of attention and the resulting competition among content gained traction as a possible explanation for the heavy-tailed popularity distributions usually observed in social media~\cite{lerman2010information}, akin to critical phenomena, which spawned a number of models~\cite{gleeson_PhysRevX.6.021019_2016, oliveira2018scaling, lorenz-spreen2019accelerating}. It remains unclear, however, how these properties relate to the other known mechanisms for the formation of information-limiting environments.

In this work, we want to understand how the spread of information affects polarization dynamics and vice-versa. In particular, we also address the gaps in the literature with respect to how to characterize information-limiting environments by investigating which mechanisms are associated with polarized states with limited spread of information.

We approach our research questions by running extensive computational experiments on an agent-based model that features agents equipped with limited memory and time-evolving opinions and connections. In addition, we implement a data calibration task on two Twitter datasets of polarized contexts, studied previously in~\cite{minici2022cascade}.




\section{Results}\label{sec2:results}

\subsection{Innovation probability introduces new states \label{sec2.1:model_characterization_results}}

We extend a previous model of social media interactions, originally proposed by de Arruda et al.~\cite{de2022modelling,de2023echo}, to explore how content sharing and opinion dynamics evolve in online social media. Our model treats opinions as continuous variables, with values between -1 and 1, influenced by interactions within an adaptive directed network, where users represent nodes and connections represent follower relationships in the social network. In this dynamic, there is a function that controls whether users post a piece of information, namely \emph{Posting}. As in~\cite{de2023echo}, we use a function of \emph{conflicting posting}, which simulates confirmation bias and user reactions (e.g., quote retweets) in social networks at the same time. The social network algorithm is then simulated in the \emph{Receiving} step, which is controlled by a parameter $\phi$. As a final step, users who disagree can rewire their connections. 

Here, we introduce a new feature of users having a memory of received posts (i.e., a social media feed), where each user maintains a list of previously received content that may influence their future posts. We also introduce an innovation probability $\mu$, which allows for the generation of new content, and probabilistic filters to simulate the effects of recommendation algorithms and user preferences on content distribution. This entire model is detailed in Section~\ref{subsec3.1:model}.

As the first step to characterize our model, we run an experiment on a known parametrization in de Arruda's model to understand the impact of the innovation probability $\mu$. To this end, we use an Erd\H{o}s-Rényi network of size $N = 10^3$ and mean in-degree $z = 10.21$, and pick parameters known to lead the system to a consensus scenario when the original model is recovered ($\mu = 1$).

\begin{figure}[H]
\centering
\includegraphics[width=0.65\linewidth]{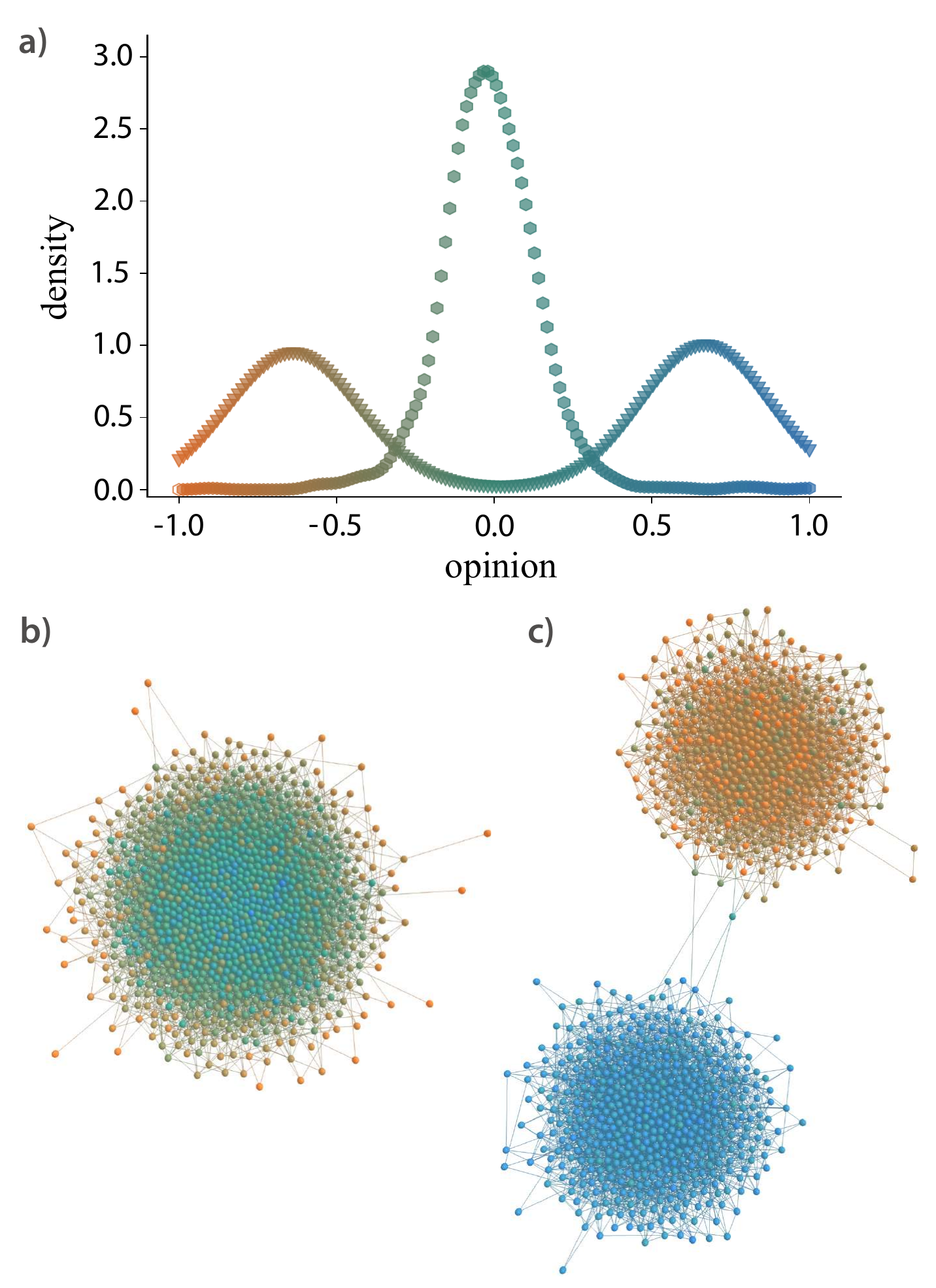}
\caption{\textbf{New system states introduced with the parameter $\mu$:} the same parameterization with the exception of different innovation probability $\mu$ can lead to consensus in a homogeneous topology or two polarized communities. In panel (a), the opinion distribution is in the other two panels, where hexagon markers represent the network in (b) and triangle markers the one in (c). In panel (b), the innovation probability $\mu=1$. In panel (c), $\mu=0.1$.}
\label{fig:multiple_states}
\end{figure}

However, running the simulation with the same parameters but a different innovation probability ($\mu=0.1$) yields a polarized system. We compare the two configurations for the same number of iterations in Figure~\ref{fig:multiple_states}. In Figure~\ref{fig:multiple_states}a, the opinion distributions are plotted against each other. We also show how each network structure is wired in each case, with nodes colored according to their opinions. Figure~\ref{fig:multiple_states}b shows the case for $\mu = 1$, which is statistically the same as the initial network, and opinions are homogeneously spread around 0. Figure~\ref{fig:multiple_states}c reveals that, for the polarized state at $\mu = 0.1$, the network becomes separated into two densely connected groups of largely opposing opinions. See Section \ref{subsec3.2:polmeasure} for more details.

In order to characterize the impact of the innovation probability $\mu$ in the opinion distribution, we run a hundred simulations on the same network for $\mu = 0$ (no innovation) and $\mu = 1$ (de Arruda's original model). We then use each of these simulations to produce a trajectory of opinion distributions with 100 points. These trajectories are then compared via the moment ratio diagram of Figure~\ref{fig:moment_ratio}, which uses L-moments~\cite{hosking2018lmoments} to describe these distributions. The L-skewness and L-kurtosis are bounded counterparts to the sample skewness and excess kurtosis. Skewness measures the asymmetry of a distribution, indicating whether the distribution of opinions is skewed. Kurtosis indicates whether the data has heavier or lighter tails compared to a normal distribution, with higher kurtosis indicating more extreme values. For more information, see Section~\ref{subsec3.2:polmeasure}. 

We show in Figure~\ref{fig:moment_ratio} how opinion distributions become very separated, even though they share the same starting point for both $\mu=0$ and $\mu=1$. This starting point was omitted to improve visibility, but it is found around ($\tau_3,\tau_4$) = (0.16,0.16), corresponding to an opinion distribution which is approximately uniform (considering the finite-sized network) between -1 and 1.

The final point of the $\mu=1$ trajectories is around (0.65,0.3), while that of the $\mu=0$ trajectories is nearly the extreme coordinates (1,1). The extreme coordinates correspond to distributions accumulated into a few specific values, as $\mu=0$ prevents the creation of posts with different opinions $\theta$ once the system is initialized. This is the same as the predominance of very few values of $\theta$ across the system (see Appendix \ref{app:extreme_moments}).

\subsection{Simulating an information-limiting environment \label{sec2.2:ILE_experiment_results}}

Next, we implement an experiment to investigate opinion dynamics parameterized to hinder the spread of information. This is done by running simulations on a network with two communities via a stochastic block model, where users in one community are initialized with opinion 1, and those in the other are initialized with opinion -1. Each community has 500 users with mean in-degree and out-degree $z=8$, and only 1\% of edges bridge the two communities. Users are forced to remain in their community as we disable the possibility of rewiring. Furthermore, we specify an alternate posting filter (Equation~\ref{eq:P_p_ali}) and compare simulation outcomes between this filter and the original one (Equation~\ref{eq:P_p_con}). See Sections \ref{subsec3.1:model} and \ref{subsec3.3:ILE_experiment_setup} for more details.

\begin{figure}[H]
\centering
\includegraphics[width=1\linewidth]{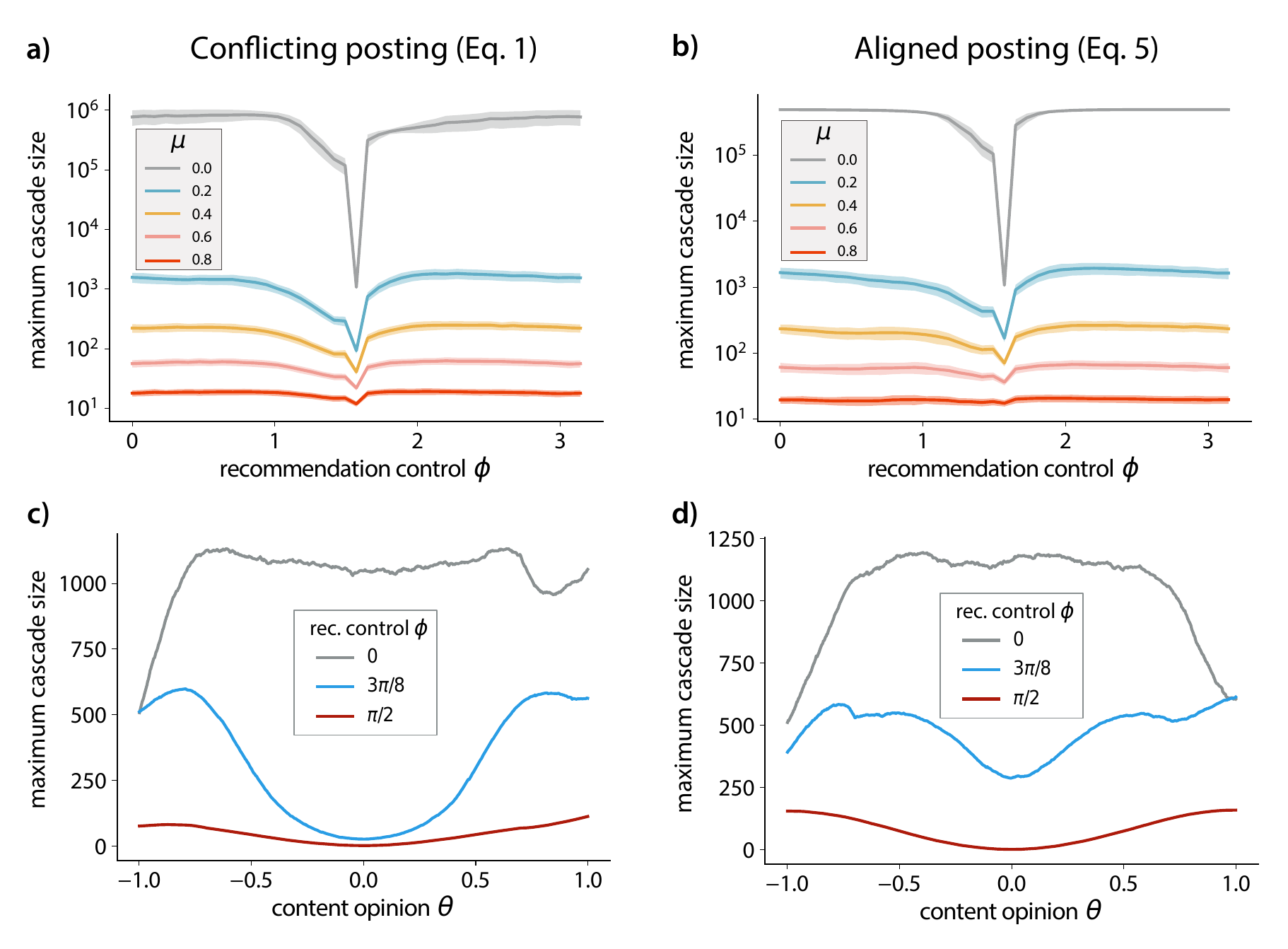}
\caption{\textbf{Effects of model calibration on information spreading in a synthetic information-limiting environment:} In panels (a) and (b), for two different posting filters, maximum cascade size produced for a range of recommendation control $\phi$ and five values of innovation probability $\mu$. For each pair of $\mu$ and $\phi$, we run 500 simulations and obtain the maximum cascade size in each of them; the solid line is the average among the 500 maximum cascade sizes, while the shade is the average added or subtracted to the corresponding standard deviation. In panels (c) and (d), again for each respective posting filter, we compare the maximum cascade size with content opinion $\theta$ for three values of recommendation control $\phi$ on a fixed innovation probability $\mu=0.2$. Since each cascade is associated with an opinion content $\theta$, which is continuous, we group cascades into each of a thousand bins between -1 and 1. From these, we extract those with the maximum cascade size; the resulting values are represented with lines smoothed via the Savitzky-Golay filter~\cite{savitzky1964smoothing}.}
\label{fig:echo_chamber_experiment}
\end{figure}

The spread of information is measured by looking at the maximum size reached by cascades, which is the same as the number of users who share the same piece of content. In Figures~\ref{fig:echo_chamber_experiment}a and \ref{fig:echo_chamber_experiment}b, we show how cascades can grow according to the relationship between the innovation probability $\mu$ and the recommendation control $\phi$, each for a specific posting filter (Equation~\ref{eq:P_p_con} for (a), Equation~\ref{eq:P_p_ali} for (b)). For each pair of $\mu$ and $\phi$, we observe the maximum cascade size in each of the 500 simulations and average them to obtain the solid line, with shades representing the difference to the corresponding standard deviation.

As expected, the lower the innovation probability $\mu$ is, the greater the sizes cascades can reach since there are fewer posts competing with each other. Both posting filters have qualitatively similar behavior, with the conflicting posting showing a slightly higher standard deviation for $\mu=0$. Our experiment setup is shown to limit the spread of information around $\phi=\pi/2$, where the receiving filter (Equation~\ref{eq:rp_I}) hinders communication within and between communities.
Next, we measure how contents spread in this scenario regarding their opinion $\theta$. By fixing $\mu = 0.2$ and three values of $\phi$, Figures~\ref{fig:echo_chamber_experiment}c and \ref{fig:echo_chamber_experiment}d show that the less permissive recommendation control regime ($\phi = \pi/2$) favors extreme content opinion, while the most permissive ($\phi = 0$) punishes it. However, the middle regime ($\phi = 3 \pi / 8$) displays different behaviors for each posting filter. In \ref{fig:echo_chamber_experiment}c, it noticeably hinders the growth of moderate content (i.e., $\theta$ around 0). In \ref{fig:echo_chamber_experiment}d, this effect is less pronounced, meaning the dynamics still permit the spread of content from the entire $\theta$ range.

\begin{figure}[H]
\centering
\includegraphics[width=0.65\linewidth]{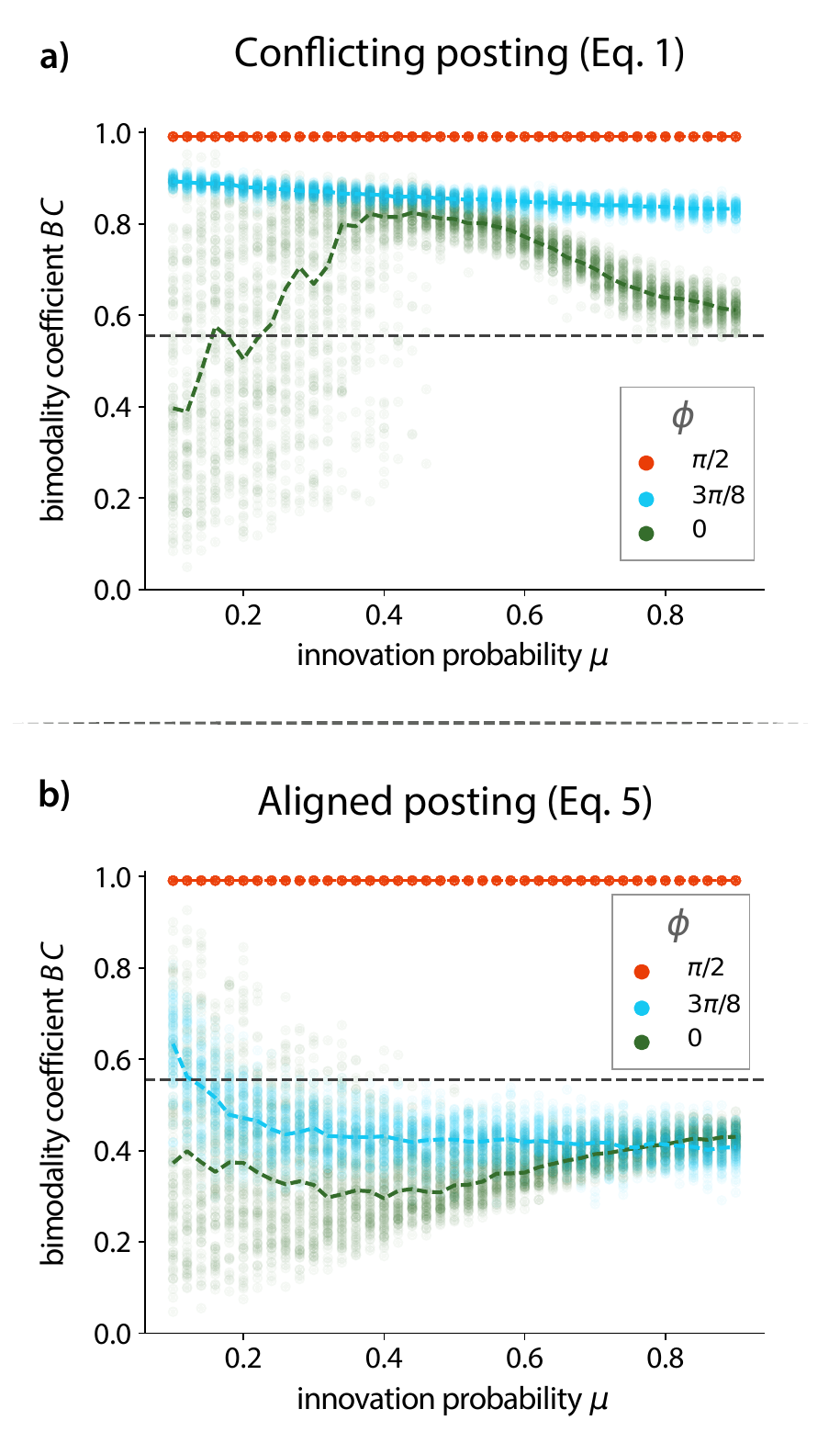}
\caption{\textbf{Polarized communities reacting to the increase of content innovation:} for each combination of posting filter (Equation~\ref{eq:P_p_con} in (a), Equation~\ref{eq:P_p_ali} in (b)), each recommendation control $\phi$ (coded by color) and each value of innovation probability $\mu$, a thousand simulations are run and represented by scatter points corresponding to the bimodality coefficient $BC$ associated to the final opinion distribution. Colored dashed lines are the median of the 1000 points at each value of $\mu$. After being initialized with extreme opposite opinions in each module, the experiment is performed without rewiring, and the resulting opinion distributions are obtained after $10^6$ iterations. The dark dashed line at $BC=5/9$ signalizes where opinion distributions are considered bimodal.}
\label{fig:depolarization}
\end{figure}

Then, we demonstrate in Figure~\ref{fig:depolarization} how the increase of content variety through the parameter $\mu$ affects the final opinion distribution for users locked into the two communities starting with extreme opinions.
To do this, we use the bimodality coefficient $BC$, defined in Section~\ref{subsec3.2:polmeasure}, which ranges from 0 to 1. Values above the threshold of 5/9 indicate bimodality, while values below it indicate unimodality.
In Figure~\ref{fig:depolarization}a, the conflicting posting filter blocks moderate posts (by not allowing their cascades to grow, as seen in Figure \ref{fig:echo_chamber_experiment}c). However, when the aligned posting filter is employed (\ref{fig:depolarization}b), the system is depolarized as the variety of content increases with $\mu$. For $\phi = \pi/2$, cascade growth is hindered, and therefore, the system remains polarized throughout the range of $\mu$ for both filters, well above the line indicating the $BC$ threshold at 5/9. But when the aligned posting filter is used, for $\phi = 3 \pi / 8$ and $\phi = 0$, we see that even though the $BC$ values of the final opinion distribution show a wide dispersion at small $\mu$, they narrow into unimodal opinion distributions as soon as $\mu=0.3$ and remain under the dashed threshold 5/9 for most of the remaining $\mu$ range.

\subsection{Model calibration against data \label{sec2.3:data_calibration_results}}

We implement a calibration task to see how our model conforms to real-world scenarios. Given the availability of network structure, user opinions, and cascade sizes in the datasets, we opt to focus on the latter as the target of our model calibration. That is, we consider the empirical cascade size distribution to be aggregated over time, while the network structure and user opinions are left as degrees of freedom in the time evolution of our model (even though all simulations start from the same network structure).

Thus, we formulate the model calibration as an optimization task to reproduce the closest possible distribution of cascade sizes over our parameter space. In particular, we search for the parameters that control the behavior of the dynamics: the innovation probability $\mu$, the recommendation control $\phi$, the opinion variation $\Delta$, and the number of iterations $n_\text{iter}$. The precise formulation of the optimization task is given in Section~\ref{subsec3.5:data_calibration_task}.

In Figure~\ref{fig:data_calibration}, we show the results for simulations parameterized according to the lowest values of KS statistic observed, which are at $\alpha = 27$ for the Brexit dataset and $\alpha = 29$ for the VaxNoVax dataset. More details regarding the optimization are shown in Supplementary Material~S1.

For both datasets, the optimized models represent most of the cascades found in the real data. Specifically, the values of the KS statistics are on average $(3.617 \pm 0.595) \times 10^{-3}$ for Brexit and $ (1.385 \pm 0.146) \times 10^{-2}$ for VaxNoVax. Note that a common choice to consider the KS statistic as a good fit is 0.05. In the case of our results, the values obtained are significantly lower. We acknowledge that there is a limitation in the tails of the distributions, which are not well represented. However, since the plot axes follow logarithmic scales, the discrepancy between data and simulation is much more pronounced at the extreme events (see Figure~\ref{fig:data_calibration}).

To further understand the opinion dynamics and the effect of the cascades on the opinions, we calculate the bimodality coefficient of the resulting opinions $BC(b)$. On average, $BC(b)$ is $(0.892 \pm 0.007)$ for Brexit and $(0.623 \pm 0.010)$ for VaxNoVax. Both values indicate that the opinion distributions are bimodal, and thus, the opinions are polarized. Interestingly, the parameters that generate cascades compatible with the real data also polarize opinions in the synthetic system. Furthermore, the bimodality of VaxNoVax is consistent with the results found in real data in our previous study~\cite{de2022modelling}, reported to be between $0.60$ and $0.67$.

The methodology used considers that the best result is not necessarily when the dynamics converge to a fixed result. However, for the best results obtained here, we found that $BC$ does not change significantly even when the simulation runs for many more iterations.

\section{Methods}\label{sec3:methods}

\subsection{Model definition} \label{subsec3.1:model}

Our model is an extension of the one introduced by de Arruda~\textit{et al.}~\cite{de2022modelling}. This model represents social media platform interactions and treats opinion as a continuous variable, bounded between -1 and 1, held by each user. Social media users are represented as nodes in an adaptive directed network. A user points to another if they receive content from them (followership relation); that is, the content is spread following the opposite direction of edges.

Each iteration of the model follows successive steps, which act as filters that determine whether a randomly activated user goes to the next step. 

In this work, we extend the model by allowing users to post previously existing content. Each piece of content has a fixed opinion value $\theta$, always drawn uniformly random upon its creation. However, here each user is equipped with a memory list of size $\alpha$, filled as they receive content from neighbors. Each user's memory can be seen as an ordered list of $\theta$ values from previous posts.

Memory lists are updated in such a way that the last piece of content received is prioritized. We index the memory list with $\theta_{c}$, such that $c \in \{ 1,...,\alpha \}$, and the lower the index, the more likely the content will be posted. New pieces of content are introduced from the beginning of the list ($\theta_{1}$) based on what users receive from their neighbors. In this case, existing pieces of content move to a lower order (i.e., $\theta_{c+1} \leftarrow \theta_{c}$).

When the list is at full capacity (there are $\alpha$ pieces of content) and new content is received, the piece at the bottom ($\theta_{\alpha}$) is removed, and all other pieces move one position lower so that the new one is placed at the top.

When reposting, users choose the first piece of content in their memory lists ($\theta_1$). Once the content is posted, it moves from the beginning to the end of the list, so it will be picked last. Hence the opinion of the content which was just posted becomes $\theta_\alpha$ in the memory list, and each other piece moves to a higher order (i.e., $\theta_{c} \leftarrow \theta_{c+1}$).

We further introduce a new parameter called the innovation probability $\mu$, which, at the posting step, controls how many iterations inject new content into the system. That is, when $\mu=1$, the base model from de Arruda~\textit{et al.}~\cite{de2022modelling} is recovered regardless of the memory lists.

Each iteration of our model is described by the following steps:

\begin{enumerate}
    \item \textit{\textbf{Activating:}} a user $i$, chosen uniformly at random, becomes active. 
    \item \textit{\textbf{Posting:}} the active user $i$ either creates a new content with probability $\mu$, or chooses the content on the top of their memory list ($\theta = \theta_1$) with probability $1-\mu$. The content is posted according to a probabilistic filter function (Equation~\ref{eq:P_p_con}) of their opinion $b_i$ and the content opinion $\theta$. 
    \item \textit{\textbf{Receiving:}} Each follower $j$ of node $i$ receives the content according to another probabilistic filter function (Equation~\ref{eq:rp_I}) of $b_j$, $b_i$ and a parameter $\phi$. When they do, the content is added to their memory list from the top. The control of this filter through $\phi$ is a proxy for a social media recommendation algorithm.
    \item \textit{\textbf{Realigning:}} Each opinion $b_j$ of followers who successfully received the content increases or decreases by a constant $\Delta$. The opinions are repelled away from the post with probability $|\theta - b_j|/2$ and attracted otherwise.
    \item \textit{\textbf{Rewiring:}} Each follower $j$ may stop following $i$ and start following another user at random through a probability proportional to the difference between $b_i$ and $b_j$. 
\end{enumerate}

In our model extension, we implement rules to define user behavior which were previously studied in \cite{de2022modelling} and \cite{de2023echo}. In particular, we set the probabilistic rule acting as the posting filter as
\begin{equation}
    P_p( \theta, b_i) = \cos^2\left( \frac{\pi}{2} |\theta - b_i| \right),
    \label{eq:P_p_con}
\end{equation}
where $\theta$ is the opinion of the content (whether created or picked from the memory list), and $b_i$ is the opinion of the active node $i$. This is referred to in \cite{de2023echo} as \textit{conflicting posting}, which is the same as $P_t^\text{pol}$ from Equation~2 in \cite{de2022modelling}.

Correspondingly, the receiving filter is given by
\begin{equation}
    P_r(b_i, b_j, \phi) = \cos^2\left( \frac{\pi}{2} |b_i - b_j| + \phi \right).
    \label{eq:rp_I}
\end{equation}
\noindent This filter was previously studied as the function $P^{\text{I}}_d$ from Equation~5 in \cite{de2022modelling}. We call the parameter $\phi$ the \textit{recommendation control}. It controls the starting point of the cosine-squared function, effectively calibrating the recommendation algorithm. That is, it may let information propagate between pairs of users who agree with each other or the opposite. Notice the receiving step does not consider the content opinion $\theta$.

\subsection{Opinion polarization measures} \label{subsec3.2:polmeasure}

The main system state we are interested in measuring is the distribution of user opinions. In particular, we characterize a state as polarized when the opinion distribution is bimodal.

To measure the bimodality of the opinion distributions, we adopt the bimodality coefficient used by Arruda~\textit{et. al}~\cite{de2022modelling}. The bimodality coefficient $BC(b)$~\cite{pfister2013good} is defined as 
\begin{equation}
    BC(b) = \frac{g^2 + 1}{k + \frac{3(n-1)^2}{(n-2)(n-3)}},
    \label{eq:BC}
\end{equation}
where $g$ is the \emph{sample skewness}~\cite{kokoska2000crc} and $k$ is the \emph{excess kurtosis}~\cite{kokoska2000crc} of the opinion distribution $b$. The distribution $b$ is typically considered bimodal when $BC(b) > 5/9$, as empirically shown in~\cite{pfister2013good}.

Another way to measure how the opinion distribution changes towards polarization is via the moment ratio diagram, which is composed of bounded versions of the third and fourth-moment ratios of the distribution called L-moments~\cite{hosking2018lmoments}. That is, given the opinion distribution $b$ and its $k$-th order statistic $b_{(k)}$, its sample $r$-th L-moment $\lambda_r$ is defined by

\begin{equation}
    \lambda_r = r^{-1} \binom{n}{r}^{-1} \sum\limits_{b_{(1)} < ... < b_{(j)} < ... b_{(r)}} (-1)^{r - j} \binom{r-1}{j} b_{(j)}.
\end{equation}

\noindent The moment ratio diagram is composed of the L-skewness $\tau_3 \coloneqq \lambda_3 / \lambda_2$, which is bounded between -1 and 1, and the L-kurtosis $\tau_4 \coloneqq \lambda_4 / \lambda_2$, which is bounded between -1/4 and 1. We choose these over the conventional moments (which are unscaled) to compare different stages in the evolution of the opinion distribution.

To produce Figures \ref{fig:multiple_states} and \ref{fig:moment_ratio}, we set up the simulation on an Erd\H{o}s-R\'enyi network of size $N=10^3$, with recommendation control $\phi = \pi/2$, opinion variation $\Delta = 0.1$ and feed size $\alpha=1$. Opinions are initialized as a uniformly random number between -1 and 1, and we execute $10^5$ iterations for Figure~\ref{fig:multiple_states}, and 100 steps of $10 N$ iterations for each value of $\mu$ in Figure~\ref{fig:moment_ratio}.

\subsection{Information-limiting environment experiment} \label{subsec3.3:ILE_experiment_setup}

Here we describe how our experiment on an information-limiting environment is designed and set. First, we produce a synthetic network with two communities separated by very few edges. We employ the Poisson degree-corrected stochastic block model as implemented in \cite{peixoto_graph-tool_2014} to generate a network with size $N=10^3$, with both in-degree and out-degree distributions from Poisson with an average $z=8$ and two modules. Only $1\%$ of the edges connect the two modules.

We want to design a polarized scenario where users are locked into two distinct communities, both separated by opinion and network structure. Hence, this experiment is executed with the rewiring rule disabled, and users have opinions initialized as 1 in one module and -1 in the other. That is, the system always starts fully polarized in this experiment.

We also highlight an important distinction to the original model setup. Instead of the posting filter from Equation~\ref{eq:P_p_con}, we opt to use the following:
\begin{equation}
    P_p( \theta, b_i )= 
\begin{cases}
    \cos^2\left( \frac{\pi}{2} |\theta - b_i| \right), & \text{if } |\theta - b_i| \leq 1\\
    0,              & \text{otherwise}.
    \label{eq:P_p_ali}
\end{cases}
\end{equation}

This change means that, in practice, a user $i$ will never post a content of opinion $\theta$ when the difference $|\theta - b_i|$ is greater than one. Since each user is initialized with an opinion of either 1 or -1, they will not post any content whose $\theta$ has a different sign to their opinion unless it has changed since the beginning of the simulation.

Then we set the opinion variation $\Delta = 0.1$ and feed size $\alpha = 1$ and proceed to understand how these features impact the growth of cascades of content posting by systematically testing how parameters react to each other, especially the recommendation control $\phi$ and innovation probability $\mu$.

To produce Figure \ref{fig:echo_chamber_experiment}a and \ref{fig:echo_chamber_experiment}b, we fix each value of $\mu$ and a range of 41 values of $\phi$ between $0$ and $\pi$. For each pair of $\mu$ and $\phi$, we run 500 simulations of $10^6$ iterations each and annotate what is the maximum cascade size resulting from each. The solid line represents the average of the 500 maximum cascade sizes, and the shades are the corresponding standard deviation added or subtracted from the average.

We obtain Figure \ref{fig:echo_chamber_experiment}c and \ref{fig:echo_chamber_experiment}d by fixing $\mu = 0.2$, three values of $\phi$ and running 100 simulations for $10^6$ iterations each. Each value of $\phi$ then generates a large number of cascades: 9,372,911 cascades for $\phi = 0$, 8,901,320 cascades for $\phi = 3 \pi / 8$ and 5,090,826 cascades for $\phi = \pi/2$. These are binned through a thousand values of content opinion $\theta$, which we then use to extract a maximum cascade size. Finally, the maximum cascade sizes for each bin of $\theta$ are represented together through smoothed lines via the Savitzky-Golay filter\cite{savitzky1964smoothing}.

Finally, Figure~\ref{fig:depolarization} considers a range of 41 values of $\mu$, from 0.1 to 0.9, and the three previously tested values of $\phi$ ($0$, $3 \pi / 8$ and $\pi / 2$). At each point given by a combination of parameters, we run 1000 simulations.

\subsection{Datasets} \label{subsec3.4:datasets}

In our study, we consider two datasets published by Minici \emph{et al.}~\cite{minici2022cascade}. They were collected from the social platform Twitter under a background of strongly polarized debates. These datasets contain measurements of network structure, cascades of retweets, and opinions of either users or posts. All of these measures are represented in our model, which makes the data good candidates for empirical studies.

\begin{figure}[H]
\centering
\includegraphics[width=0.65\linewidth]{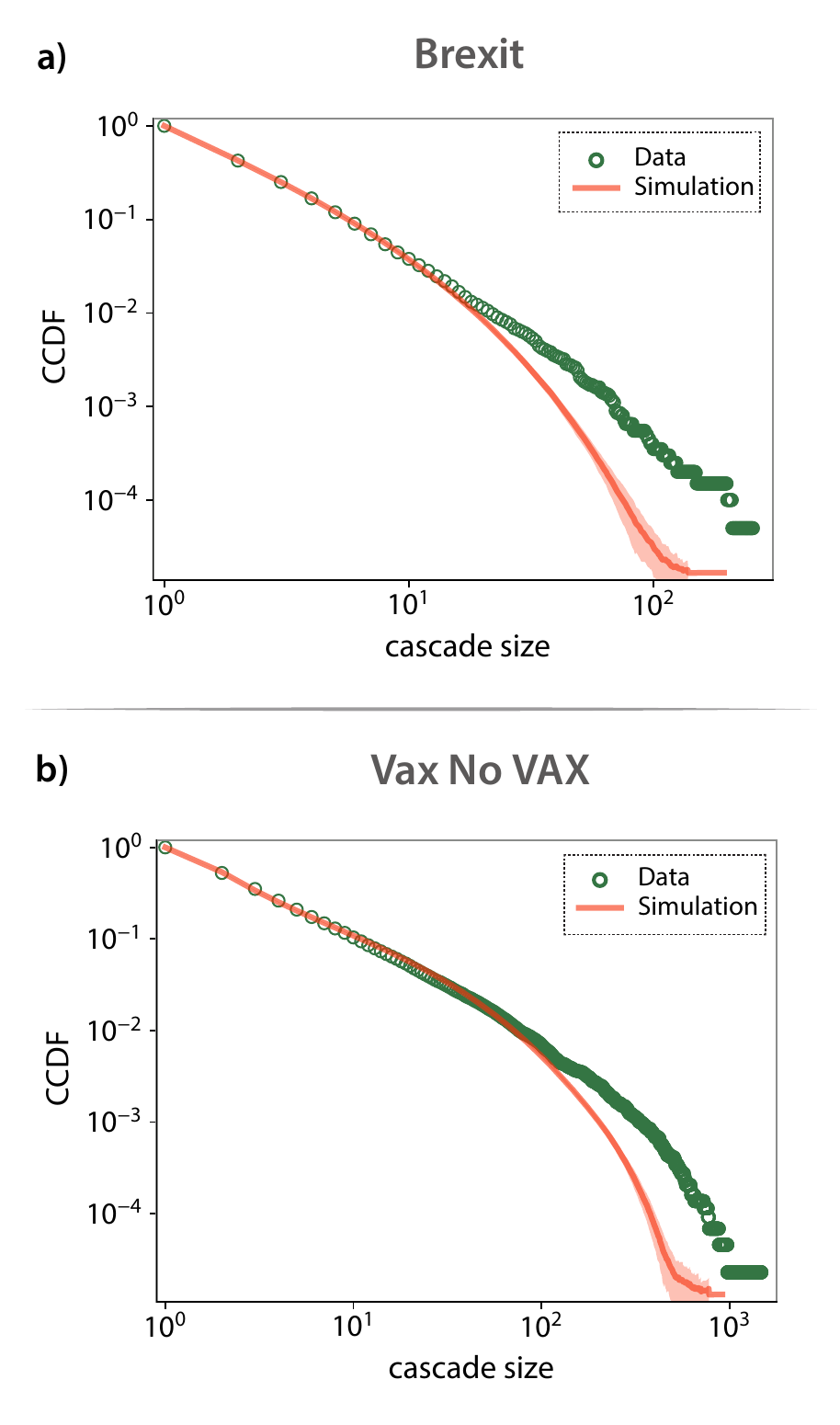}
\caption{\textbf{Calibrating the model against data:} complementary cumulative distribution function (CCDF) of cascade sizes obtained from the optimal simulation against those from each dataset. The green line corresponds to the average of 100 simulations, while the shades account for the standard deviation of simulated cascade sizes. In panel (a), the model is calibrated against cascades from the Brexit dataset, and the set of optimal simulation parameters found is $(\mu,\Delta,\phi,n_\text{iter}) = (0.658, 0.373, 1.366, 6.379 \times 10^5)$ for a feed size $\alpha = 27$. The KS statistic between the sample distribution from data and those from simulations averages at $(3.617 \pm 0.595) \times 10^{-3}$. In panel (b), the model is calibrated against cascades from the VaxNoVax dataset, and the set of optimal parameters is $(\mu,\Delta,\phi,n_\text{iter}) = (0.093, 0.724, 2.863, 6.474 \times 10^5)$ for a feed size $\alpha = 29$. The KS statistic between the sample distribution from data and those from simulations averages at $(1.385 \pm 0.146) \times 10^{-2}$.}
\label{fig:data_calibration}
\end{figure}

The first dataset, named `Brexit', was measured in the context of the Brexit referendum in the UK between May and July of 2016. It features a network with 7,589 nodes and 532,460 edges. The number of cascades is 19,963, with minimum and maximum sizes of 2 and 256, respectively. The complementary cumulative distribution function (CCDF) of the cascade sizes is shown in Figure~\ref{fig:data_calibration}a.

The second dataset, dubbed `VaxNoVax', was collected during vaccine debates in Italy in 2018. Its associated network has 14,315 nodes and 1,714,180 edges. It contains 43,923 cascades ranging in size from 2 to 1,468 (see Figure~\ref{fig:data_calibration}b). 
We also provide the degree distribution of both networks in the Appendix \ref{app:datasets}.

\subsection{Model calibration task} \label{subsec3.5:data_calibration_task}
Our model is calibrated via a search in a multidimensional space made of four parameters, which are $\mu$, $\phi$, $\Delta$, and $n_\text{iter}$. In particular, we minimize the Kolmogorov-Smirnov (KS) distance via a single-objective genetic algorithm from the implementation in~\cite{pymoo} by considering the feed size $\alpha$ as a fixed parameter when the heuristic is initialized. The values of $\alpha$ are selected considering the KS statistic obtained for a range up to $\alpha = 45$, found in the Supplementary Material.

Since the datasets are a small sample \cite{morstatter2021sample} of the activity in subsets of the Twitter platform retrieved by the first version of the API, it is a challenging task to accurately estimate the parameters \textit{a priori} due to their dependence on time. We cannot know how the volume of activity represented in the sample maps to the actual one in the system at the time. In essence, it is not trivial to match one simulation iteration to a time scale in the real world.

As such, two of our parameters encapsulate the time frame we are trying to match: the number of iterations $n_\text{iter}$, but also the opinion variation $\Delta$ as it modulates how intense are changes in user opinion. Intuitively, a high value of $\Delta$ means fewer iterations are needed for the system to reach stable configurations. 
Once there is a time window set by $n_\text{iter}$ and $\Delta$, the innovation probability $\mu$ is fundamental to fix the maximum cascade size. 

Finally, the recommendation control $\phi$, mentioned back in Section~\ref{subsec3.1:model} as related to how users receive posts of different opinion, can drive the system towards various states of opinion polarization (as observed in previous works~\cite{de2022modelling,de2023echo}).

Let $d$ be the empirical distribution of cascade sizes from data and $s$ the sample distribution of cascade sizes obtained from the simulation. As each dataset has a fixed distribution of cascade sizes, let the empirical cumulative distribution function of $d$ be given by $F_d (x)$. As for the sample distribution $s$, it is parameterized with the innovation probability $\mu$, the recommendation control $\phi$, the opinion variation $\Delta$, and the number of iterations $n_\text{iter}$.

We are then interested in finding the simulation parameters that minimize the Kolmogorov-Smirnov (KS) statistic~\cite{Corder_Foreman_2014} between the two samples. That is, the model calibration task is
\begin{mini!}|l|
{\mu,\phi,\Delta,n_\text{iter}}{\sup_{x}|F_{d}(x) - F_{s}(x,\mu,\phi,\Delta,\alpha,n_\text{iter})|}
{}{}
\addConstraint{0.01 \leq \mu \leq 0.99}{}
\addConstraint{0 \leq \phi \leq \pi}{}
\addConstraint{0 \leq \Delta \leq 1}{}
\addConstraint{1.5 \times 10^5 \leq n_\text{iter} \leq 6.5 \times 10^5}{}
\end{mini!}


\noindent where $F_{s}(x,\mu,\Delta,\phi,\alpha,n_\text{iter})$ is the empirical cumulative distribution of cascade sizes from simulations parameterized accordingly.

After fixing a value of $\alpha$, which we obtain by testing the KS statistic for different ranges of parameters (see Appendix~\ref{app:opt}), we obtain an optimal set of parameters with the genetic algorithm. From this set of parameters, Figure~\ref{fig:data_calibration} is produced by comparing the empirical cascade size tail distribution in the datasets to the one produced by averaging 100 simulations with the set of parameters.

These simulations are executed using the empirical network structure from the dataset, with opinions initialized uniformly at random between the entire interval [-1,1]. Each iteration follows the description of Section~\ref{subsec3.1:model}, including the posting filter given by Equation~\ref{eq:P_p_con}.








\section{Discussion}\label{sec12}

In this work, we introduced an agent-based model for opinion polarization on adaptive networks featuring a limited-attention mechanism. The latter allows the same content to be reposted, which in turn means we can keep track of cascades of information formed by reposting from different users. We studied the behavior of this model to understand the interplay between the spread of information and opinion polarization dynamics.

We first characterized how the innovation probability $\mu$ changes a previously known scenario on an Erdős-Rényi network by observing the evolution of opinions. Then, we investigated the model behavior in a scenario designed to hinder the spread of information. In this scenario, a synthetic modular network locks users into one of two communities initialized with opposing extreme opinions, with only 1\% of the edges between them. From this experiment, we saw it is possible to go from polarization to consensus depending on the model parameters and posting filters, even when users cannot rewire.

Finally, we applied the intuition gained from these experiments to the study of empirical data by calibrating our model so that it produced cascades of posts with similar statistics to those observed in the real world. Two datasets with empirical network structure and cascade statistics, from the platform previously known as Twitter were used to evaluate how well the parameter space adapts.



Even though at least 99\% of cascade sizes empirically observed are well captured within the optimal set of parameters, we opted to showcase the results through a log-log plot, as it is customary for analyzing distributions of critical phenomena. This visualization option shifts the focus away from the vast majority of events and towards the extreme events, which is also consistent with the previous analyses using the cascade maximum size.

The deviations between the tails of the two distributions may be attributed to a number of different factors. Here we highlight that our agent-based iterations oversimplify the temporal dynamics involved in the system we study by considering the homogeneous activity, while extreme events in social media platforms are often modeled with bursty dynamics (\cite{Karsai2017Bursty},\cite{Notarmuzi2022universality}).

Furthermore, our data calibration task was a very specific attempt at joining computational experiments and empirical phenomena. Extensive fitting tasks would require a bigger number of datasets with observations of various statistics so that different objective functions can be set up.

Another measure that can be readily compared in the calibration task is the ratio between the number of distinct cascades over the sum of all their sizes, which works an upper bound for the empirical innovation probability (see Equation 53 of \cite{gleeson_PhysRevX.6.021019_2016}). In the case of the VaxNoVax dataset, this bound is $\Tilde{\mu} = 0.1444$, which is compatible with the optimal value of innovation probability retrieved by our task ($\mu = 0.093$).

As per the Brexit dataset, $\Tilde{\mu} = 0.272$ does not match the optimal parameter $\mu = 0.658$. Here it is important to highlight that the Brexit dataset may pose limitations to our investigation. The authors in \cite{minici2022cascade} mention that this particular dataset was collected in two steps since the first did not contain retweets. After the second collection, they filtered out all posts that did not contain at least one of the 100 most popular hashtags, as well as users who had only four or fewer tweets. This means the retweet cascades may be incomplete to some degree, which influences all the other correlated measures.

While we build on the understanding of previous investigations, which guide a lot of our choice of parameters and filter functions, the present work is no longer a simple iteration of the related work. Our model is a novel framework to study the temporal evolution of opinion phenomena produced by the conjoined effects of several features of social media platforms, such as the effect of content recommendation, connection rewiring, and limited user attention.

Among the vast range of possible configurations and their outcomes, we focused our contribution on polarized settings and which mechanisms in these settings may hinder the spread of information. Future endeavors may turn to alternate configurations or even dig further into additional explanation candidates for polarized scenarios, such as rewiring bias or recommendation bias.

Allowing the individual tracking of posts cascading through the network on top of the adaptive opinion dynamics contributes to new tools for investigating mechanisms in social media dynamics. Though it is interesting to aggregate statistics per cascade, another possible approach is to aggregate cascade statistics on the user level. That could allow for explaining how mediators between two communities behave and whether or how they control the flow of information between two sides.

\backmatter





\bmhead{Acknowledgements}
Y.M was partially supported by the Government of Aragón, Spain and ``ERDF A way of making Europe'' through grant E36-23R (FENOL), and by Ministerio de Ciencia e Innovación, Agencia Española de Investigación 
(MCIN/AEI/ 10.13039/501100011033) Grant No. PID2020-115800GB-I00. We acknowledge the use of the computational resources of COSNET Lab at Institute BIFI, which was funded by Banco Santander (grant Santander-UZ 2020/0274) and the Government of Aragón (grant UZ-164255). The funders had no role in study design, data collection and analysis, decision to publish, or preparation of the manuscript.

\section*{Declarations}
Not applicable

\bibliography{sn-bibliography}%









\begin{figure}[H]
\centering
\includegraphics[width=0.65\linewidth]{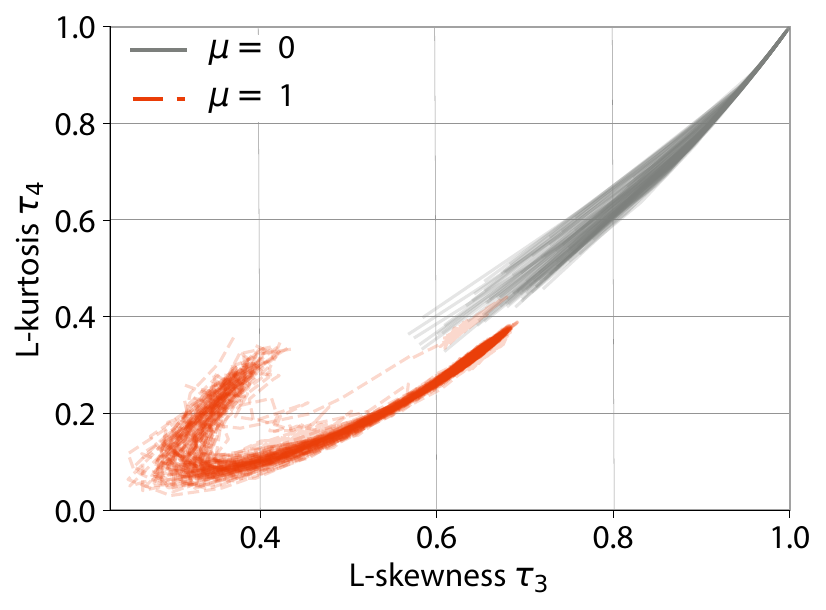}
\caption{\textbf{Characterizing the effect of innovation on opinion evolution:} the moment ratio diagram for the evolution of the opinion distribution in an Erd\H{o}s-R\'enyi network of size $N=10^3$ and mean in-degree $z=10.21$ for the minimum and maximum innovation probability $\mu$. Both parameterizations are tested with 100 different trajectories starting from the same initial opinion distribution (uniformly random between -1 and 1), omitted in the diagram for clarity. Each trajectory has 100 points, and every two consecutive points are produced from steps of $10N$ iterations of the model.}
\label{fig:moment_ratio}
\end{figure}

\newpage
\begin{appendices}

\section{Extreme opinion distributions for no innovation \label{app:extreme_moments}}

Here we sample one final opinion distribution found at the maximum L-skewness $\tau_3$ and L-kurtosis $\tau_4$ in the moment ratio diagram of Figure \ref{fig:moment_ratio} and show how several predominant opinions from independent trajectories are distributed.

\begin{figure*}[h]
\includegraphics[width=1\linewidth]{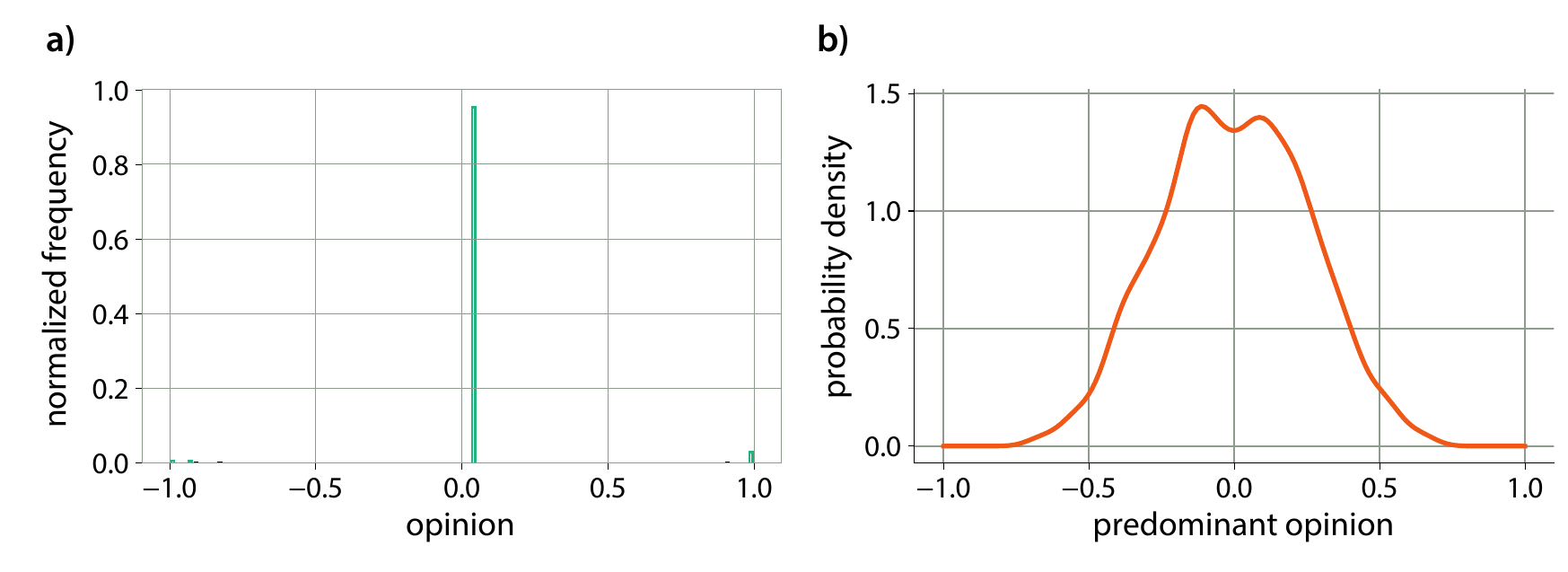}
\caption{In (a), a sample distribution corresponding to maximum L-skewness $\tau_3$ and L-kurtosis $\tau_4$. Considering the most prevalent value of opinion (near 0) as the predominant opinion, in (b), we show how predominant opinions extracted from $10^4$ independent trajectories are distributed.}
\label{fig:degr_distrs}
\end{figure*}




\section{Dataset additional characterization \label{app:datasets}}

In Figure \ref{fig:degr_distrs}, we show the tail distributions for in-degrees and out-degrees of both datasets published by Minici \emph{et al.}~\cite{minici2022cascade}, available at \url{https://github.com/mminici/Echo-Chamber-Detection/tree/main/data/raw}.

\begin{figure*}[h]
\includegraphics[width=1\linewidth]{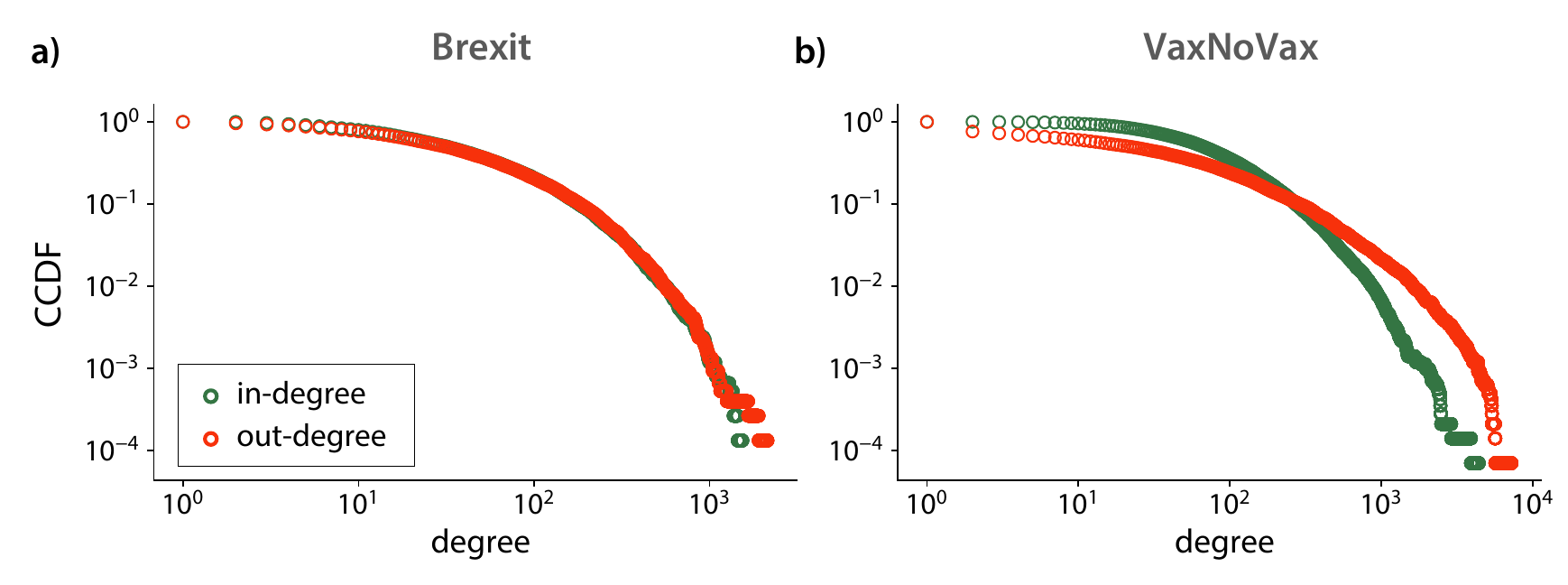}
\caption{Complementary cumulative distribution function (CCDF), or tail distribution of degrees from the empirical networks used in our model calibration task. The data was made available by \cite{minici2022cascade}.}
\label{fig:degr_distrs}
\end{figure*}

\section{Optimization of real-world data \label{app:opt}}
\begin{figure}[h!]
\includegraphics[width=1\linewidth]{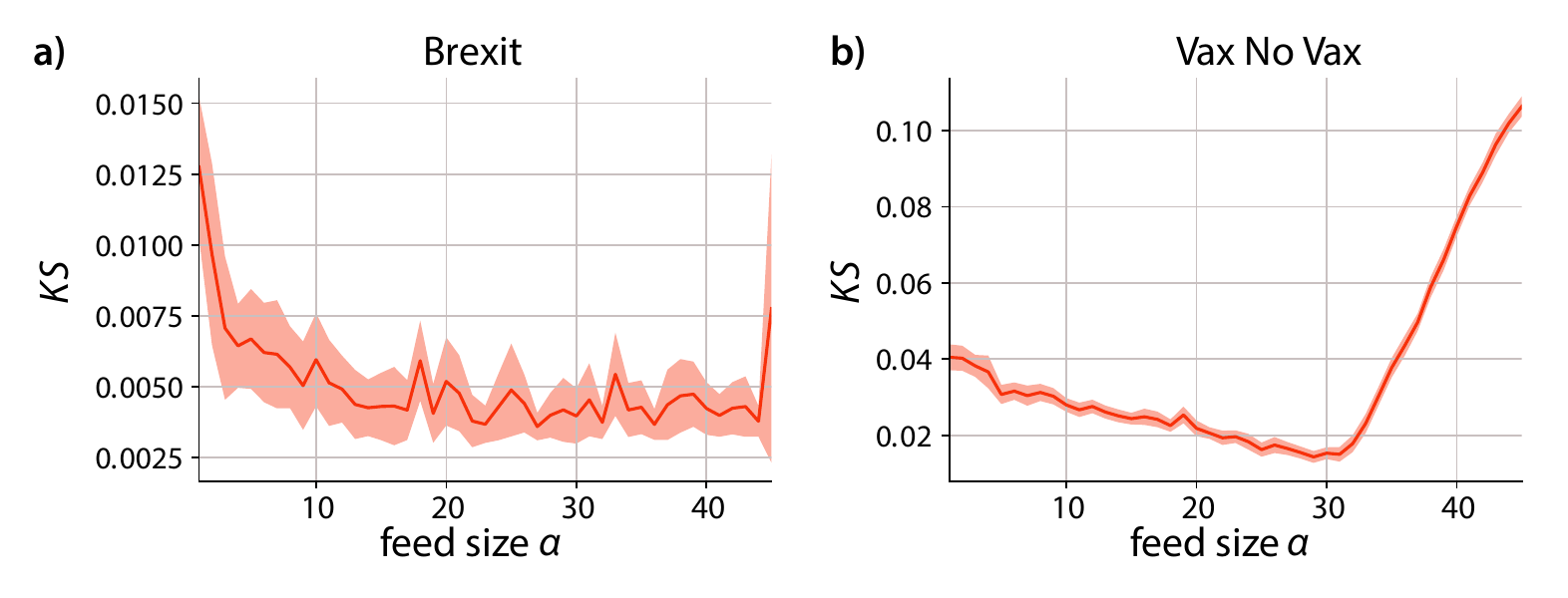}
\caption{Average and standard deviation of the KS statistic between the real data and the data generated by the opinion model obtained for the optimized parameters, for each value of $\alpha$. To generate this plot, we run the dynamics 300 times for each value of $\alpha$.}
\label{fig:optimization_alphas}
\end{figure}

As described in the main text, we optimize the opinion dynamics with respect to the parameters $\mu$, $\phi$, $\Delta$, $\alpha$, and $n_\text{iter}$ using a genetic algorithm. However, to reduce the computation time and understand the effect of the feed size ($\alpha$), we do not consider it a parameter to optimize. The dynamics are run once for each parameter set tested to reduce the optimization time further. We understand that the results obtained from this approach can be penalized by some statistical fluctuations. Still, we argue that if the optimizer slightly changes a parameter, the dynamics is rerun. Thus, we expect that the fluctuations are compensated for by running the model with parameters close to the expected minimum. Furthermore, there is no guarantee that the obtained result is the global minimum, but we show that our approach is working correctly when we see the values of the KS statistic, which are considerably low. Figure~\ref{fig:optimization_alphas} shows the best KS statistic for the parameters obtained.

\begin{figure*}[h!]
\includegraphics[width=1\linewidth]{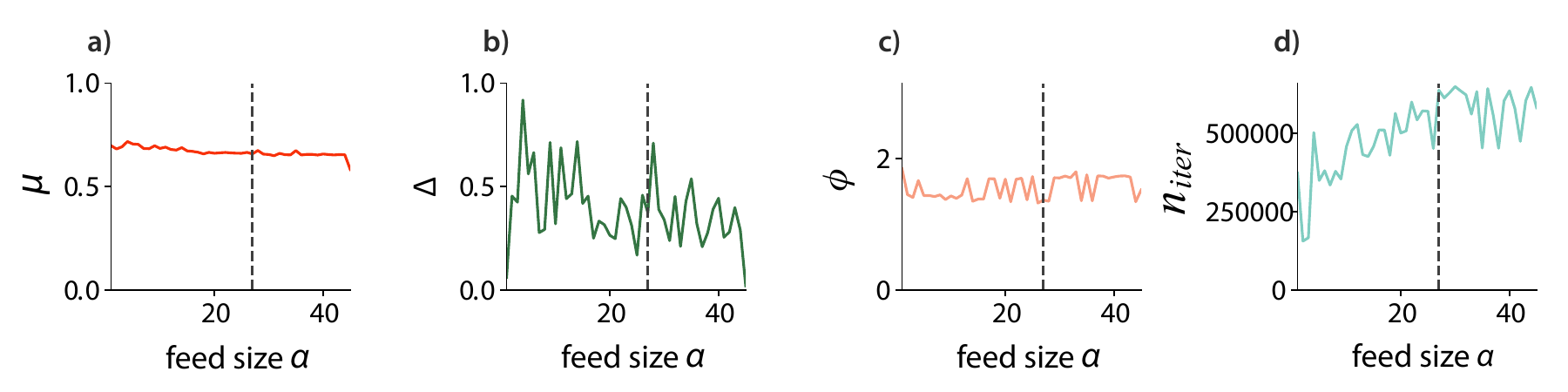}
\caption{Parameters found by optimizing the dynamics for each value of $\alpha$ for the Brexit dataset. The dashed line represents $\alpha=27$, which matches the minimum value obtained for the KS statistic.}
\label{fig:optimization_parameters_brexit}
\end{figure*}

\begin{figure*}[h!]
\includegraphics[width=1\linewidth]{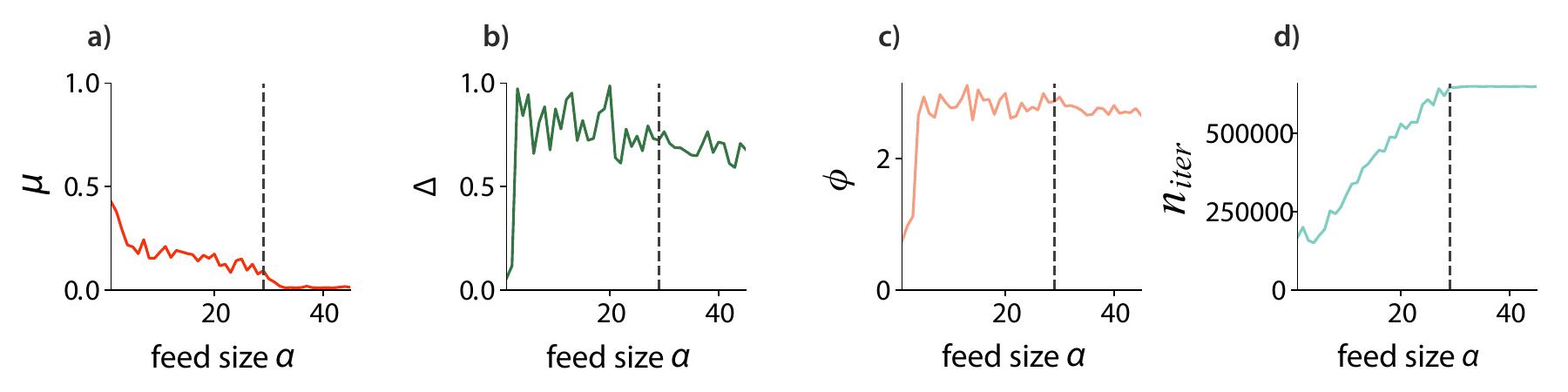}
\caption{Parameters found by optimizing the dynamics for each value of $\alpha$ for the VaxNoVax data set. The dashed line represents $\alpha=29$, which matches the minimum value obtained for the KS statistic.}
\label{fig:optimization_parameters_vax}
\end{figure*}

As a complement, Figures~\ref{fig:optimization_parameters_brexit}~and~\ref{fig:optimization_parameters_vax} show the parameters obtained from the optimization process for the Brexit and VaxNoVax datasets, respectively, varying $\alpha$. In the case of Brexit, $\mu$ and $\phi$ are more stable. In contrast, $\Delta$ and $n_{\text{iter}}$ are less stable. These results indicate that the innovation and configuration of the receiving filter tend to be the same for different feed sizes, regardless of the feed size. In the case of VaxNoVax, only the parameter $\phi$ is stable regardless of the feed size. Thus, the configuration of the receiving filter does not vary significantly as a result of $\alpha$. Interestingly, from the $\alpha$ obtained from the best KS statistics, $n_{\text{iter}}$ stabilizes in the case of VaxNoVax, but this effect does not happen for Brexit.

\end{appendices}



\end{document}